\documentclass[fleqn,usenatbib]{mnras}

\usepackage{newtxtext,newtxmath}

\usepackage[T1]{fontenc}

\DeclareRobustCommand{\VAN}[3]{#2}
\let\VANthebibliography\thebibliography
\def\thebibliography{\DeclareRobustCommand{\VAN}[3]{##3}\VANthebibliography}

\usepackage{graphicx}	
\usepackage{amsmath}	

\title[Accreting sdOB-WD binary candidates]{Limiting the accretion disk light in two mass transferring hot subdwarf binaries}

\author[Deshmukh et al.]{Kunal Deshmukh,$^{1}$\thanks{E-mail: kunaldes225@gmail.com}
Thomas Kupfer,$^{1}$
Pasi Hakala,$^{2}$
Evan B. Bauer,$^{3}$
Andrei Berdyugin,$^{4}$
Lars Bildsten,$^{5,6}$ \newauthor
Thomas R. Marsh,$^{7}$
Sandro Mereghetti,$^{8}$
Vilppu Piirola$^{4}$
\\
$^{1}$Department of Physics and Astronomy, Texas Tech University, PO Box 41051, Lubbock, TX 79409, USA \\
$^{2}$Finnish Centre for Astronomy with ESO (FINCA), Quantum, FI-20014 University of Turku, Finland\\
$^{3}$Center for Astrophysics | Harvard \& Smithsonian, 60 Garden St, Cambridge, MA 02138, USA\\
$^{4}$Department of Physics and Astronomy, 20014 University of Turku, Finland\\
$^{5}$Department of Physics, University of California, Santa Barbara, CA 93106, USA\\
$^{6}$Kavli Institute for Theoretical Physics, University of California, Santa Barbara, CA 93106, USA\\
$^{7}$Department of Physics, University of Warwick, Coventry CV4 7AL, UK\\
$^{8}$INAF – Istituto di Astrofisica Spaziale e Fisica Cosmica, Via A. Corti 12, I-20133 Milano, Italy\\
}

\date{Accepted 2022 November 21. Received 2022 November 19; in original form 2022 September 9}

\pubyear{2022}

\begin{document}
\label{firstpage}
\pagerange{\pageref{firstpage}--\pageref{lastpage}}
\maketitle

\begin{abstract}
We report the results from follow-up observations of two Roche-lobe filling hot subdwarf binaries with white dwarf companions predicted to have accretion disks. ZTF\,J213056.71+442046.5 (ZTF\,J2130) with a 39-minute period and ZTF\,J205515.98+465106.5 (ZTF\,J2055) with a 56-minute period were both discovered as subdwarf binaries with light curves that could only be explained well by including an accretion disk in their models. We performed a detailed high-resolution spectral analysis using Keck/ESI to search for possible accretion features for both objects. We also employed polarimetric analysis using the Nordic Optical Telescope (NOT) for ZTF\,J2130. We did not find any signatures of an accretion disk in either object, and placed upper limits on the flux contribution and variation in degree of polarisation due to the disk. Owing to the short 39-minute period and availability of photometric data over six years for ZTF\,J2130, we conducted an extensive $O - C$ timing analysis in an attempt to look for orbital decay due to gravitational wave radiation. No such decay was detected conclusively, and a few more years of data paired with precise and consistent timing measurements were deemed necessary to constrain $\dot P$ observationally.

\end{abstract}




\section{Introduction}

Subdwarf O/B (sdOB) stars are spectral type O/B stars with much lower luminosities than the main sequence, and appear on or near the extreme horizontal branch (EHB) in the Hertzsprung-Russell diagram. Most sdOBs are believed to be helium core burning stars which lost their hydrogen rich envelope \citep{heb86,heb09,heb16}. Although the exact formation channel is still not fully understood, it has been shown that binary evolution plays a significant role and may even be required to form sdOBs \citep{nap04a,max01,pel20}. Systems with orbital periods below a few days are formed through a common envelope (CE) phase which is followed by the loss of angular momentum due to the radiation of gravitational waves \citep{han02,han03,nel10a}. The most compact sdOB binaries have orbital periods below 1\,hr \citep{kup20,kup20a}. 

SdOB binaries with white dwarf (WD) companions which exit the CE at an orbital $\leq$2\,hrs will start mass transfer to the WD companion while the sdOB is still burning helium. When the binary leaves the CE, gravitational wave radiation carries away angular momentum from the binary. This shrinks the orbit until the sdOB fills its Roche Lobe at a period of $\approx30-100$\,min depending on the evolutionary stage of the sdOB (e.g. \citealt{sav86,tut89,tut90,ibe91,yun08,pie14,bro15,neu19,bau21}). Mass transfer of helium rich material from an sdOB to a WD companion could either disrupt the WD even when the mass is significantly below the Chandrasekhar mass, a so-called double detonation supernova  (e.g. \citealt{liv90,liv95,fin10,woo11,wan12,she14,wan18}) or just detonate the He-shell without disrupting the WD which results in a faint and fast .Ia supernova with subsequent weaker He-flashes \citep{bil07,bro15}. 

The known population of sdOB binaries with orbital periods below 2\,hours is low. Only five detached systems with a WD companion are known to have orbital periods below 2\,hours \citep{ven12,gei13,kup17,kup17a, pel21, kup22}. Recently \citet{kup20,kup20a} discovered the first sdOB+WD binaries with likely ongoing accretion. Both systems were discovered as part of a high cadence low Galactic latitude survey by the Zwicky Transient Facility \citep[ZTF]{kup21}. ZTF\,J213056.71+442046.5 (ZTF\,J2130) with a 39-min period and ZTF\,J205515.98+465106.5 (ZTF\,J2055) with a 56-min period have light curves that could not be explained with ellipsoidal modulation of the sdOB star alone. Modifying the light curve models of the binaries by adding an edge-irradiated accretion disk component which eclipses the sdOB donor star resulted in excellent fits to the data. The light curve shape is still the sole hint towards an accretion disk and theoretical models predict an accretion rate that could lead to $\lesssim$1\% flux contribution from the disk (\citealt{kup20,kup20a}). 

The spectroscopic observations reported in both discovery papers were mostly done in low resolution and did not report accretion disk spectral signatures. \cite{RiveraSandoval2019} reported on a 1 ks X-ray observation of ZTF\,J2130 by the Neil Gehrels Swift Observatory. This was followed by a much deeper 65 ks observation by XMM-Newton. Both observations were unable to detect any X-ray flux from the system. Based on the latter observation, \cite{Mereghetti2022} placed an upper limit of $0.5-2.5\times10^{30}$ erg/s for the luminosity, with the exact value depending on the assumed spectral shape. ZTF\,J2130 is expected to be a strong gravitational wave source and \citet{kup20} estimated based on the system properties an orbital decay of $\dot P=(-1.68\pm0.42) \times 10^{-12}s \text{ }s^{-1}$. They predicted that this should be detectable with eclipse timing measurements after a few years of photometric monitoring. 

In this paper we present results from follow-up observations using high-resolution spectroscopy as well as spectral polarimetry to search for direct evidence for an accretion disk. Additionally, we present results based on timing data for ZTF\,J2130 spanning over six years. The layout for this paper goes as follows - Section~\ref{section:obs} describes all the observations categorised by spectroscopy, polarimetry and photometry. Section~\ref{section:spec} describes the spectral analysis of ZTF\,J2130 and ZTF\,J2055, followed by Section~\ref{section:pol} on polarimetric analysis of ZTF\,J2130. We discuss timing analysis of ZTF\,J2130 in Section~\ref{section:time} and end with conclusions in Section~\ref{section:con}.

\section{Observations}
\label{section:obs}

\subsection{Spectroscopy}

ZTF\,J2130 and ZTF\,J2055 were observed using Keck with the Echellette Spectrograph and Imager (ESI) in Echelle mode resulting in a resolution of $R=6000$ and wavelength coverage over 3919 - 10145 Angstrom. The data were obtained on two consecutive nights in July 2020. A total of 110 phase resolved spectra with an exposure time of 180\,s covering seven orbits for ZTF\,J2055; and 87 spectra with an exposure time of 120\,s covering six orbits for ZTF\,J2130 were obtained. They were reduced using the MAKEE\footnote{\url{https://sites.astro.caltech.edu/~tb/makee/}} pipeline, which comprises of bias subtraction, flat fielding, sky subtraction, order extraction, and wavelength calibration. This data was used for spectral analysis, particularly to look for possible emission or absorption features from the accretion disk.

\subsection{Polarimetry}

ZTF\,J2130 was observed at the Nordic Optical Telescope (NOT) during four consecutive nights on 20th - 24th 
of July, 2020 with the DiPol-UF three-bands polarimeter \citep{piir}. Altogether 2h+2h+5h+5h of data were obtained during those nights. We used 3-second exposure times, which provided a polarimetric time resolution of $\sim$15 seconds. A total of 3150 polarimetric measurements were obtained. For linear polarization measurements, a super-achromatic half-wave plate (HWP) is used as polarization modulator. A plane-parallel calcite plate placed below the HWP divides the light beam into two parallel orthogonally polarised beams which, after the wavelength separation made by two dichroic beam-splitters, are registered simultaneously and independently by three EM CCD cameras. We emphasize that with DiPol-UF the blue, visual and red passbands ($B^{\prime},V^{\prime}, R^{\prime}$) are not identical to those of the Johnson -- Cousins system, but are defined by the dichroic beam-splitters and additional sharp cut-off filters with high peak transmission. See details and transmission curves in \citet{berd2022}. Four consecutive exposures, taken at wave-plate orientations separated by the angle of 22.5 degrees, yield single measurement of Stokes parameters $q$ and $u$. See \citet{piir} for the detailed description of the instrument and the data reductions. Each exposure shows dual (ordinary and extraordinary beam) images of the source and the nearby comparison star located at 13” in South-South West (SSW) direction of the source.

Standard CCD image calibration methods (dark, bias and flat-fielding) have been employed prior to extraction of
the o- and e-images with the aperture photometry algorithm. Stokes parameters $q$ and $u$ are then determined from the o- and e-image intensity ratios. Those Stokes parameters have been corrected for the telescope polarization and transformed from the instrumental to the equatorial celestial coordinate system. For the determination of the telescope polarization, we have observed nearby non-polarised stars HD 136064, HD 152598, and HD 159332, from the list published by \citet{piir2020}. The telescope polarization was found to be $< 0.01\%$ in all passbands and thus negligible for the present case. For the determination of the polarization angle zero-point, two highly-polarised standard stars HD 161056 and HD 204827 have been observed.

The obtained polarization data were also useful for timing analysis of brightness variations. We used only the fourth night's $B^{\prime}$-band data, which had 4516 points of 3 second exposure each. Times were originally in the start-of-exposure Heliocentric Julian Date (HJD) format. We converted these to mid-exposure Modified Julian Dates (MJDs), and then to the corresponding Barycentric Julian Dates (BJDs), as was done for all photometric observations described in the next subsections.

\begin{table*}
	\centering
	\caption{Summary of all observations used in this work}
	\label{tab:obs}
	\begin{tabular}{llllll} 
		\hline
		Date & Telescope/Instrument & N$_{\text{exp}}$ & Exp. time (s) & Coverage(A$^0$)/Filter \\
		\hline
		\textbf{Spectroscopy - ZTF\,J2055}  \\
		2020-07-21 - 2020-07-22 & Keck/ESI & 110 & 180 & 3919 - 10145\\
		\hline
		\textbf{Spectroscopy - ZTF\,J2130 } \\
		2020-07-21 - 2020-07-22 & Keck/ESI & 87 & 120 & 3919 - 10145\\
		\hline
		\textbf{Polarimetry - ZTF\,J2130 } \\
		2020-07-21 - 2020-07-24 & NOT & 12596 & 3 & $B^{\prime},V^{\prime}, R^{\prime}$\\
		\hline
		\textbf{Photometry - ZTF\,J2130 }  \\
		2018-12-13 & Palomar 48 inch & 269 & 30 &  ZTF-r \\
		2019-07-08 & GTC/HiPERCAM & 1576 & 1.77 &  g$_{s}$ \\
		2019-08-15 - 2019-08-22 & TESS & 5000 & 120 & 6000-10000 \\
		2022-09-02 - 2022-09-07 & TESS & 20000 & 20 & 6000-10000 \\
		2020-07-24 & NOT & 4516 & 3 & B\\
		2021-01-06 - 2021-01-07 & XMM-Newton/OM & 2640 & 10 &  UVW1 \\
		2021-09-04 & McDonald/ProEM & 635 & 5 &  g \\
		2021-09-06 & McDonald/ProEM & 666 & 5 &  g \\
		2021-10-12 & McDonald/ProEM & 1057 & 5  & g \\
		2022-01-08 & McDonald/ProEM & 592 & 5  & g \\
		2021-01-09 & McDonald/ProEM & 533 & 5 & g \\
		2015-07-29 - 2022-01-15 & ATLAS & 1596 & 30 &  o-band\\
		\hline
	\end{tabular}
\end{table*}

\subsection{Photometry}

Extensive timing data for ZTF\,J2130 were obtained using pointed observations from multiple telescopes as well as from surveys. To have all times in a common format, we concluded upon the mid-exposure BJD as the appropriate format. For data in other formats, we used astropy.time\footnote{\url{https://docs.astropy.org/en/stable/time/index.html}} \citep{astpy13, astpy18} for conversion to corresponding BJDs. Following are the brief descriptions of all data used for timing analysis.

\subsubsection{ZTF}
ZTF\,J2130 was observed by ZTF in 2018 as part of the high-cadence ZTF Galactic Plane survey in the r-band \citep{kup20,gra19,bel19}. Image processing of ZTF data is described in full detail in \citet{mas19}. The object was observed over two consecutive nights for 3 hours and a little under 3 hours respectively, with a total 537 data points of 30 second exposure each. Although there are other observations in both ZTF-$r$ and ZTF-$g$ scattered over a few years, we have only used the high-cadence ZTF-$r$ light curve for the purpose of timing analysis. 

The data were obtained from the ZTF/IRSA interface \footnote{\url{https://irsa.ipac.caltech.edu/Missions/ztf.html}}. The timestamps were originally in the Modified Julian Date (MJD) format for the start of each observation. These times were off-set by 15 seconds to get mid-exposure MJDs, and then converted to mid-exposure BJDs.

\subsubsection{GTC/HiPERCAM}
The HiPERCAM instrument mounted on Gran Telescopio Canarias (GTC) observed ZTF\,J2130 in multiple bands simultaneously for a duration of 46 minutes, a little over one whole period of the binary \citep{dhi16,dhi18,dhi21}. We use the same g-band data for our analysis as presented in \citet{kup20}. Owing to an exposure time of just 1.77 seconds and a total of 1576 data points, these data were the most precise for timing measurements. The default timestamps were mid-exposure MJDs, and were subsequently converted to mid-exposure BJDs.

\subsubsection{XMM-Newton}
Pointed X-ray observations of the binary were conducted using XMM-Newton in January 2021 \citep{Mereghetti2022}. The Optical Monitor (OM) instrument onboard XMM-Newton simultaneously obtained UV observations over multiple orbits in UVW1 (291 nm) and UVW2 (212 nm) bands \citep{mas01}. Since the UVW1 filter had a much higher count rate, and hence a much better signal-to-noise ratio (SNR), we found these data to be more suitable for timing analysis. ZTF\,J2130 was observed for 26.4 ks in UVW1. We used these data to extract a light curve binned at 10\,s, with mid-exposure times in BJDs.

\subsubsection{McDonald}
The most recent observations of the binary have been made using the 2.1 m Otto Struve Telescope at McDonald Observatory. We used g-band data from five different nights spanning from September 2021 to January 2022. Each observation covered one to two orbits, giving a few hundred data points of 5 second exposure each. All times were obtained as mid-exposure MJDs, and were converted to corresponding mid-exposure BJDs.

\subsubsection{TESS}
ZTF\,J2130 was in the field of view of the Transiting Exoplanet Survey Satellite (TESS) telescope \citep{ric15} once from August 2019 through October 2019 (Sectors 15 and 16), and again in September 2022 (Sector 56). The data from 2019 were taken at a 2 minute cadence, considerably longer than other telescopes. However, this data covered over a thousand orbits of the system and would be computationally expensive to analyse. In order to optimise our timing analysis, we limited our selection to a subset of 5000 points while still including around 250 orbits of the binary. On the other hand, data from 2022 were available at a 20 second cadence. Owing to the better cadence, we selected a subset of 20000 points covering only around 170 orbits. Data were obtained from the Mikulski Archive for Space Telescopes (MAST) interface and all timestamps were already in the required mid-exposure BJD format.

\subsubsection{ATLAS}
ATLAS is a survey project meant for detecting asteroids, and scans the whole sky using four telescopes, two in Hawaii, one each in Chile and South Africa \citep{ton18,hei18}. Although there were no high cadence data for the binary in this survey, it was observed several hundred times over the last seven years. We used the Forced Photometry online tool provided by the survey to obtain data from as early as 2015. Observations were taken in the o-band (orange) and c-band (cyan), of which we used only the former due to higher cadence. Data with high errors in photometry were filtered out. Times corresponding to start-of-exposure MJDs were recorded, which we converted to the desired mid-exposure BJD format.

The observations used for this paper are summarised in table~\ref{tab:obs}.

\section{Spectral analysis}
\label{section:spec}

Preliminary spectral analysis for ZTF\,J2130 \citep{kup20} and ZTF\,J2055 \citep{kup20a} was discussed in the respective discovery papers which focused on the measurement of the binary parameters. Further high resolution spectra observations by Keck/ESI were motivated by the indication towards the presence of an accretion disk from the light curve modelling. As noted in Table~\ref{tab:obs}, a total of 87 two-minute long spectra were obtained for ZTF\,J2130, and 110 three-minute long spectra for ZTF\,J2055. Since the sdOB is the donor in both cases, we expect the composition to be similar in the accretion disk. The sdOB surface is dominated by hydrogen and helium, so a signature from the disk is most likely to be associated with the corresponding spectral lines. Moreover, this signature, if present, would have radial velocities 180 degrees out of phase with the sdOB spectral lines because the accretion disk would follow the motion of the accreting white dwarf companion. We expect an emission feature since the disk is irradiated by the sdOB, however our method applies equally well for a possible absorption feature. By visual inspection of the spectra in the vicinity of the sdOB H and He lines as well as otherwise, we did not find any obvious pattern corresponding to out-of-phase sinusoidal variation.

In order to detect accretion features, it would be ideal to have good phase resolution as well as a high SNR. First, we phase folded the data over their respective orbital periods. To improve the SNR while also maintaining a decent phase resolution, we compromised by binning the data into 10 uniformly spaced phase bins. We weighted each data point by its fraction that overlaps with a bin. For example, a data point lying completely inside a bin was given a weight of 1 for that bin and 0 for all other bins, whereas a data point lying $x \%$ in one bin and $y \%$ in a neighbouring bin was given weights of $x/100$ and $y/100$ for the respective bins and 0 for all other bins. The exposure time for both objects corresponded to less than 0.1 phase units, so no data point overlapped with more than two bins. After calculating weights by this method, we coadded the data points to obtain phase-resolved spectra with, on average, nine spectra per phase bin for ZTF\,J2130 and eleven for ZTF\,J2055. These correspond to SNR improvements of $\sim$3 and $\sim$3.3 respectively. For the remainder of the section, we refer to these bins as the spectra. The spectra for both objects were analysed similarly, and we describe our analysis in the following paragraphs.

We determined the radial velocities for the spectra using sdOB hydrogen and helium lines. The radial velocity semi-amplitudes for ZTF\,J2130 and ZTF\,J2055 were consistent with \citet{kup20} and \citet{kup20a} respectively, with slightly larger error bars due to binning. For each source, the 10 phase-binned spectra were shifted to zero radial velocity and coadded to obtain a "master" spectrum with an even higher SNR. This master spectrum was then shifted to the radial velocity of each spectrum individually and used to divide it and obtain a ratio spectrum. This was done to essentially get rid of all sdOB spectral features, which simply cancelled out on taking the ratio. On the other hand, since any feature from the disk would be out of phase with the sdOB, we expect it to stand out and show up in a sinusoidal phase-dependent pattern. We looked for such a pattern all over the ratio spectra, particularly in the proximity of sdOB spectral lines. No clear pattern was found for either object through visual inspection.

\begin{figure*}
	\includegraphics[width=\textwidth]{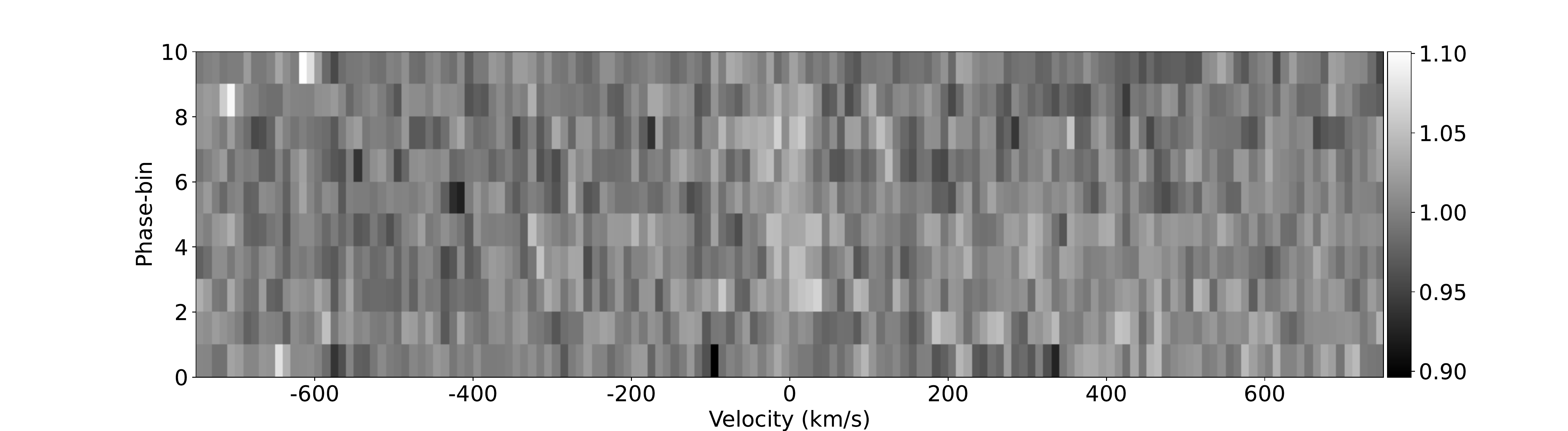}
    \caption{ZTF\,J2055 ratio velocity spectra obtained from coadding H$_{\beta}$, H$_{\gamma}$, HeII 4686 and H$_{\alpha}$ lines, shown as a heatmap plot.}
    \label{fig:j2055}
\end{figure*}

\begin{figure*}
	\includegraphics[width=\textwidth]{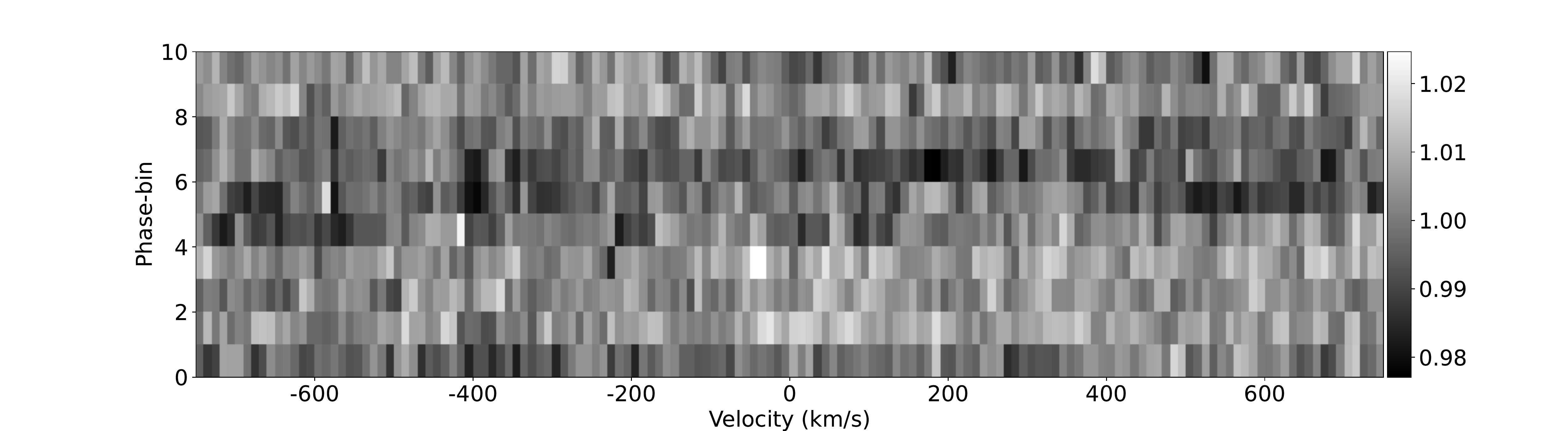}
    \caption{ZTF\,J2130 ratio velocity spectra obtained from coadding H$_{\beta}$, H$_{\gamma}$, HeII 4686 and H$_{\alpha}$ lines, shown as a heatmap plot.}
    \label{fig:j2130}
\end{figure*}

The rotation and surface gravity of the sdOB is mainly responsible for the width of its spectral lines. However, this width also depends on the wavelength, making it vary for different spectral lines. This is quantified by the relation $\Delta v = c\Delta \lambda/ \lambda$. Since all sdOB spectral lines correspond to a common $\Delta v$, going from wavelength space to velocity space would facilitate further coaddition to improve SNR.  We took each ratio spectrum and coadded multiple regions centered around typically observed spectral lines (H$_{\beta}$, H$_{\gamma}$, HeII 4686 and H$_{\alpha}$) in the velocity space. This was done by converting a short interval of data in the neighbourhood of these lines to velocity space using $\Delta v = c\Delta \lambda/ \lambda$. This procedure was followed to obtain ten "velocity" spectra. Similar to previous analysis, a master velocity spectrum was used to divide all the velocity spectra to finally obtain ratio velocity spectra. These are shown in the form of heatmap plots in figures~\ref{fig:j2055} and ~\ref{fig:j2130}. No phase-dependent sinusoidal pattern from a possible accretion disk feature was seen in the ratio velocity spectra for either binary. 

We performed a standard deviation analysis on the ratio velocity spectra to constrain the flux contribution of the disk. A 3-$\sigma$ upper limit of 6.8\% for ZTF\,J2055 and 2.0\% for ZTF\,J2130 was obtained on the accretion disk contribution to the total flux.

\section{Polarimetric Analysis of ZTF\,J2130}
\label{section:pol}

In an accretion disk, Thomson scattering is the dominant effect which can lead to a detectable component of linear polarization in the total light of the system. The seed photons for Thomson scattering would originate from the sdOB star and possibly also from the disk itself. The polarisation is expected to vary with the orbital period \citep{bro78}. 

Thomson scattering can also occur in the atmosphere of the donor sdOB star. In this case, a polarisation signal could be expected if part of the sdOB is obscured by the optically thick accretion disk around phase 0.5 (superior conjunction of the sdOB) breaking the symmetry of visible scattering directions. A short review on polarization in binary stars is presented in \citet{pii10}.

\begin{figure*}
	\includegraphics[]{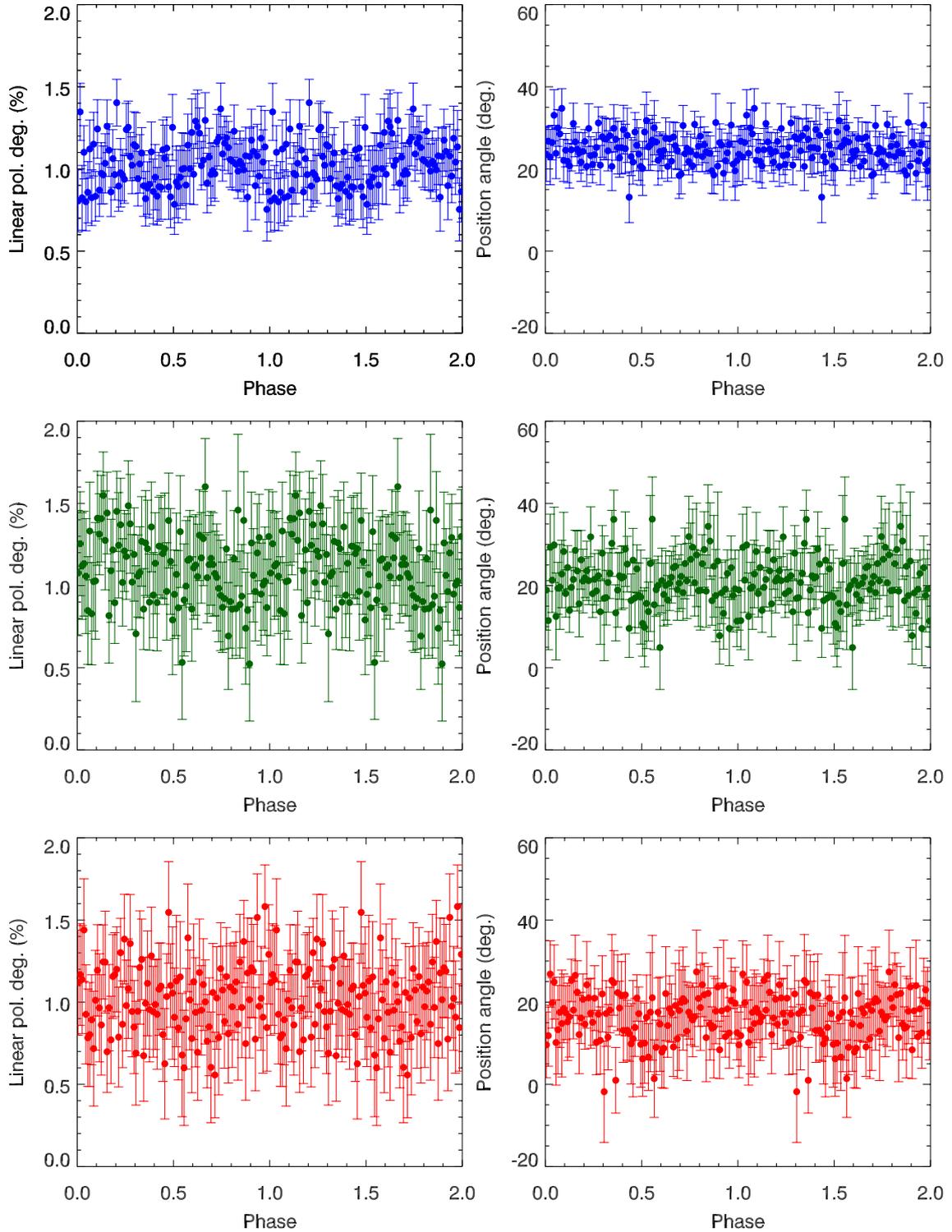}
    \caption{NOT/DiPol-UF $B^{\prime}V^{\prime}R^{\prime}$ polarimetry of ZTF\,J2130. The colours correspond to BVR bands (top to bottom). The data are shown twice for clarity.}
    \label{fig:notpol}
\end{figure*}

We do not detect any significantly variable linear polarisation from ZTF\,J2130 over the orbital period. The $B^{\prime}, V^{\prime}$ and $R^{\prime}$ polarisation curves are plotted in Fig \ref{fig:notpol}. We have estimated the upper limit for the polarisation modulation amplitude in each band by bootstrapping. We took the phase folded Q and U curves and produced 100000 artificial Q and U curves (assuming no modulation) by resampling the 100 phase bins in randomised order. We then fitted the resulting bootstrapped Q and U curves with a second order Fourier series (often used for modelling polarisation modulation by scattering in binaries \citep{bro78}) and then computed the degree of polarisation from the Q and U model fits. We then measured the 4-$\sigma$ upper limit for the variation amplitude of the degree of polarisation over the orbital period (i.e. 99.94\% of the fits having amplitude less than this value). We find 0.10\%, 0.12\% and 0.12\% 4-$\sigma$ amplitude upper limits for $B^{\prime}, V^{\prime}$ and $R^{\prime}$ respectively.      

The mean degree of polarisation (and its position angle) are very close to the values of the field star of similar brightness about 13” SSW of the target. This strongly suggests that the measured polarisation is of interstellar origin and not intrinsic to the source.

\section{Timing Analysis of ZTF\,J2130}
\label{section:time}

ZTF\,J2130 is expected to be undergoing orbital decay by losing angular momentum due to gravitational wave radiation. \cite{kup20} calculated a $\dot P=(-1.68\pm0.42) \times 10^{-12}s \text{ }s^{-1}$ using known system parameters and assuming gravitational wave radiation to be the only reason for orbital decay. This would lead to a shift of a few seconds in the expected eclipse times over a timespan of a few years. We attempted to obtain a $\dot P$ through direct timing analysis of the photometric data listed in table~\ref{tab:obs}.

For this purpose, we take the $O - C$ analysis approach. The secondary mid-eclipse, when the disk is eclipsed by the sdOB, is used as the reference point ($T_{0}$) for each orbit. We fit a model to each light curve using the LCURVE code \citep{cop10} and determine the $T_{0}$ value. In order to do so, we used the same model parameters as \citet{kup20} and froze them all with the exception of $T_{0}$. However, since we have data from different filters, we adjusted the limb and gravity darkening coefficients accordingly using values from \citet{cla11}. Since we do not have those values for XMM-Newton (UV) and ATLAS (o-band) data, we used the corresponding g-band values. While these coefficients might slightly affect the shape of the light curve model, we do not expect them to affect the mid-eclipse times. 

Assuming we know the precise orbital period, knowing one reference mid-eclipse time (call it $E_{0}$) should allow us to predict the $E^{\text{th}}$ mid-eclipse in the past as well as the future with respect to $E_{0}$. However, if the orbital period changes over time, the observed $T_{0}$'s (O) will be different from the calculated $T_{0}$'s (C). As described by \citet{Kepler1991}, we can use a Taylor series expansion to get  -

\begin{equation}
    \label{eq:o-c}
    O - C = \Delta E_{0} + \Delta P_{0} E + \frac{1}{2}P_{0}\dot P E^{2} + ...
\end{equation}

where $\Delta E_{0}$ is the error for the reference mid-eclipse time $E_{0}$, $\Delta P_{0}$ is the error in the period $P_{0}$ and $E$ is the eclipse number with respect to $E_{0}$.

\begin{figure}
	\includegraphics[width=\columnwidth]{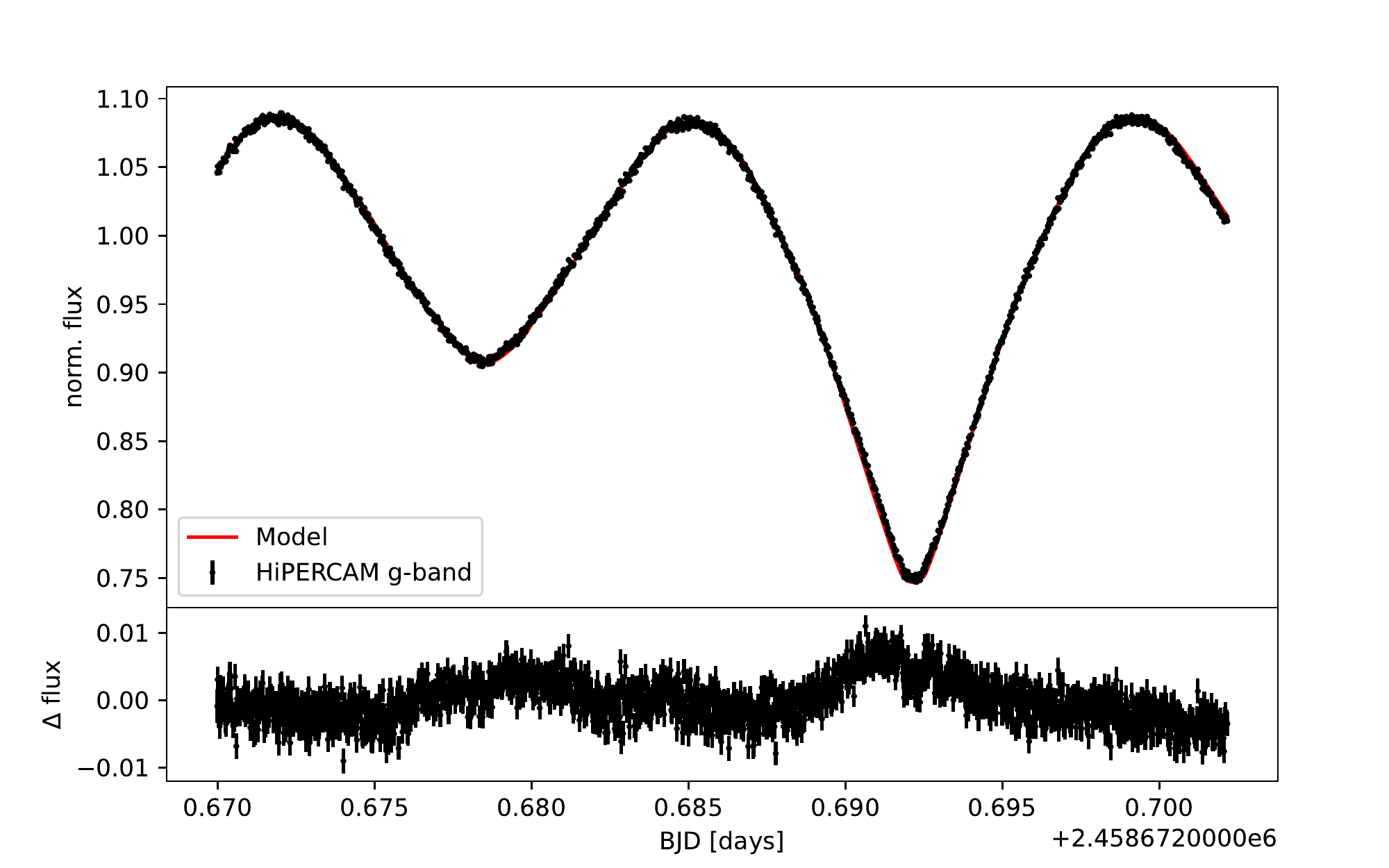}
    \caption{Best fit LCURVE model (red curve) for HiPERCAM g-band data (black points) for ZTF\,J2130. Residuals are shown in the lower panel.}
    \label{fig:hicam}
\end{figure}

\begin{figure}
	\includegraphics[width=\columnwidth]{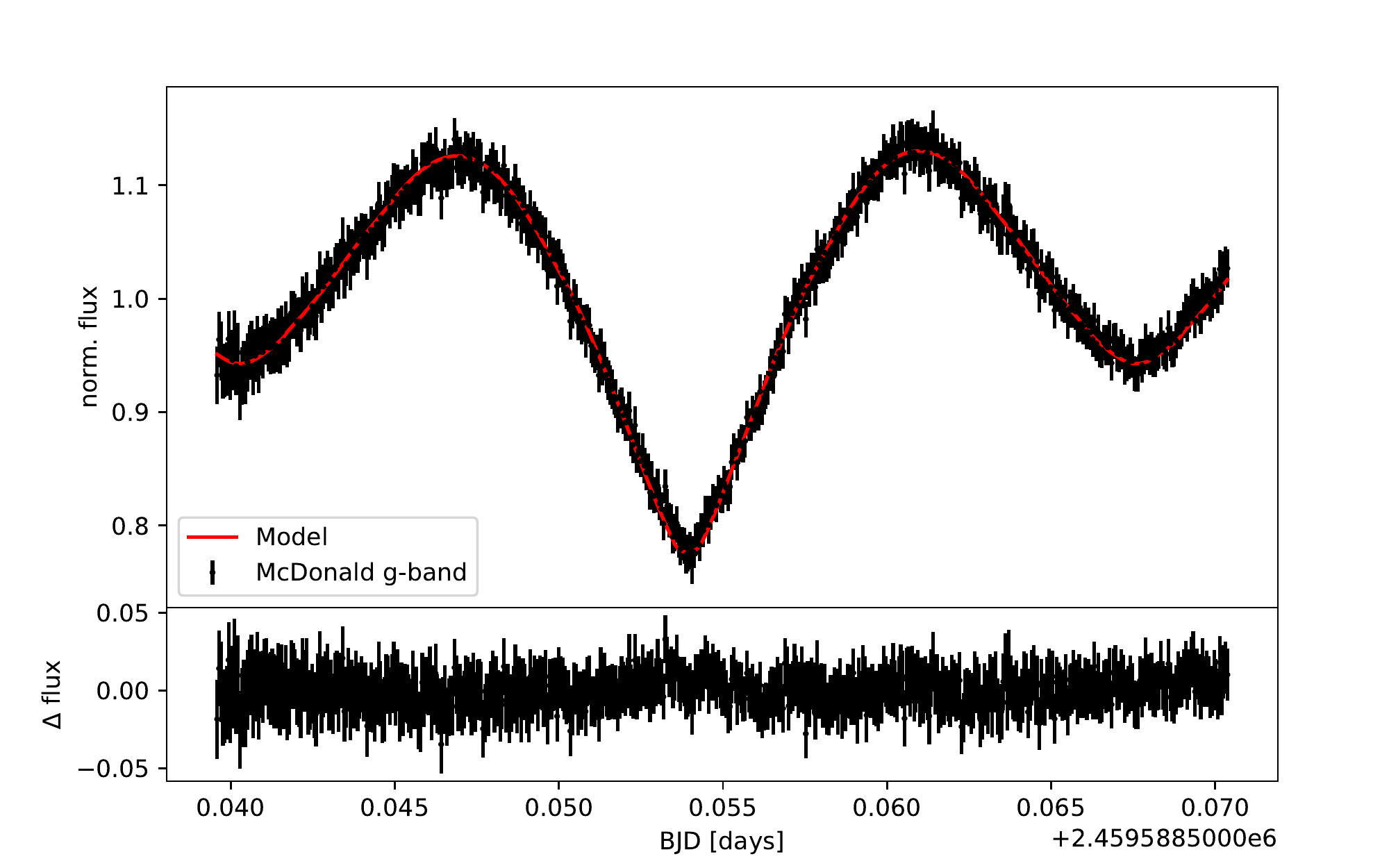}
    \caption{Best fit LCURVE model (red curve) for McDonald 2.1m g-band data (black points) for ZTF\,J2130. Residuals are shown in the lower panel.}
    \label{fig:mcd}
\end{figure}

\begin{figure*}
	\includegraphics[width=\textwidth]{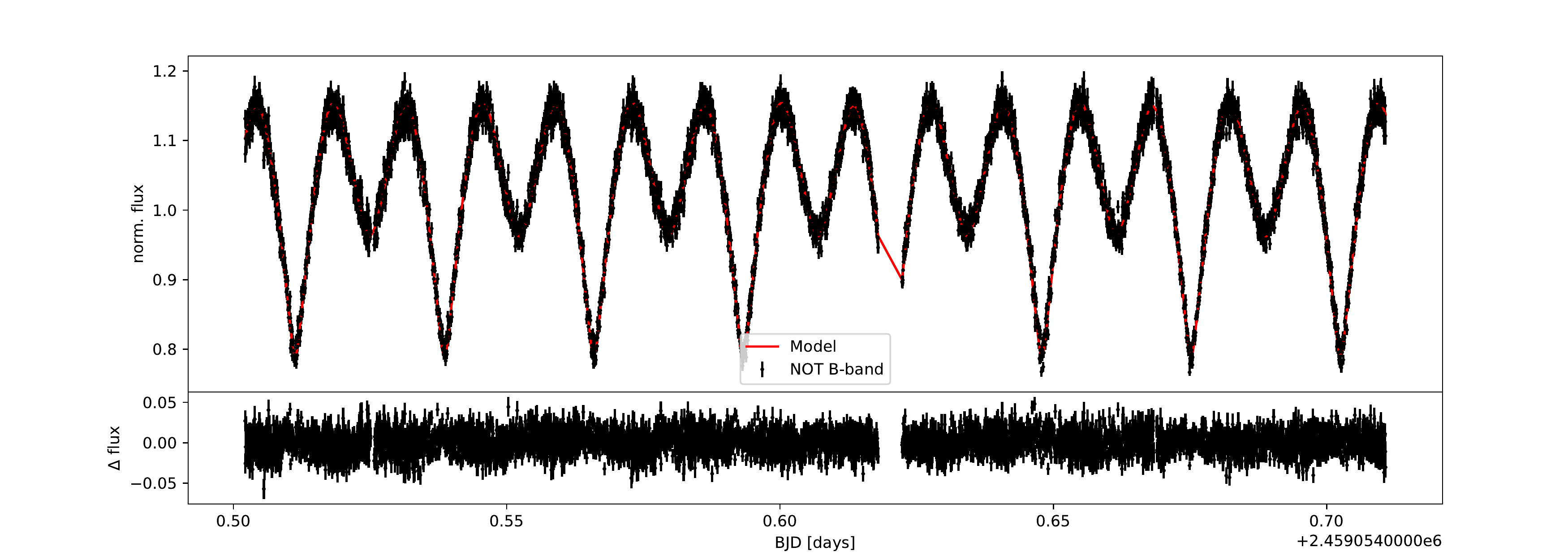}
    \caption{Best fit LCURVE model (red curve) for NOT B-band data (black points) for ZTF\,J2130. Residuals are shown in the lower panel.}
    \label{fig:not}
\end{figure*}

\begin{figure*}
	\includegraphics[width=\textwidth]{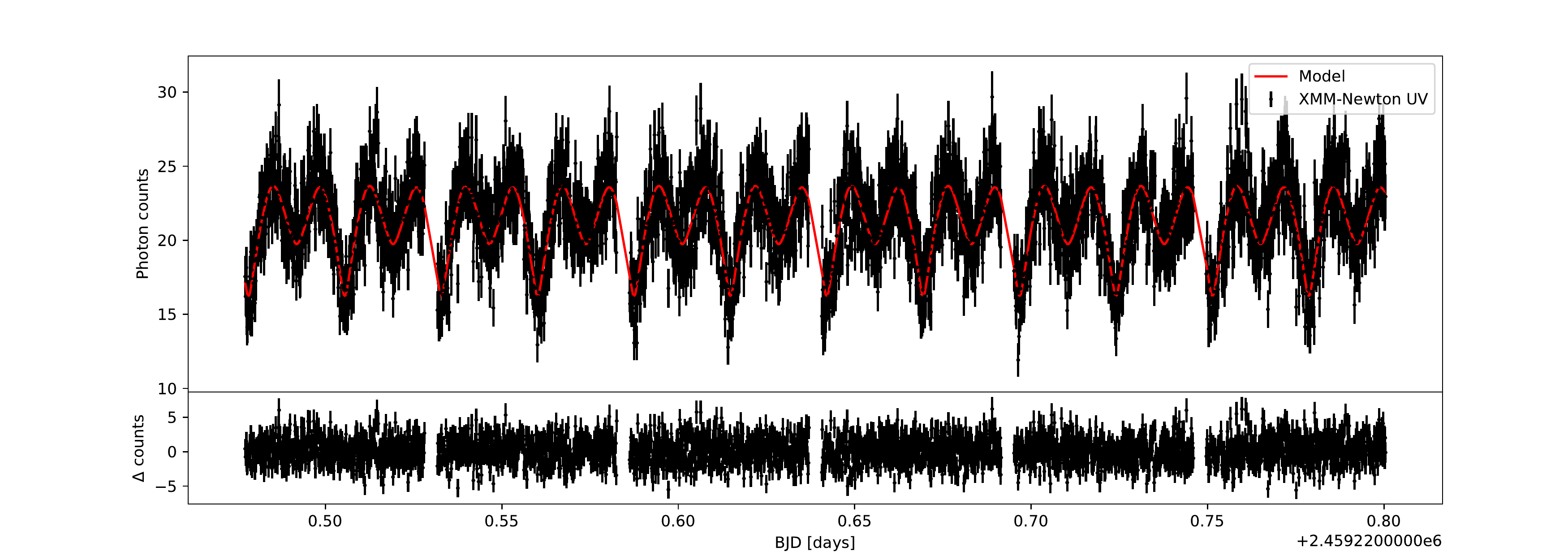}
    \caption{Best fit LCURVE model (red curve) for XMM-Newton UV data (black points) for ZTF\,J2130. Residuals are shown in the lower panel.}
    \label{fig:xmm}
\end{figure*}

\begin{table}
	\centering
	\caption{Final $T_{0}$, error in $T_{0}$ and $O - C$ values obtained for all observations}
	\label{tab:o-c}
	\begin{tabular}{llrr} 
		\hline
		Telescope & $T_{0}$ (BJD) & $O - C$ (s) & Error (s)  \\
		\hline
		ATLAS & 2457648.93660646 & -2.4 & 3.7 \\
		ATLAS & 2458440.65608749 & -10.5 & 1.9 \\
		ZTF & 2458465.59913867 & 18.3 & 2.3 \\
		HiPERCAM & 2458672.68085911 & 0 & 0.1 \\
		ATLAS & 2458674.62063972 & 8.2 & 1.9 \\
		TESS & 2458711.39258980 & -2.0 & 5.7 \\
		NOT & 2459054.52572931 & -0.5 & 0.3 \\
		ATLAS & 2459104.19255986 &  -4.8 & 1.2 \\
		XMM-Newton & 2459220.49185533 & 5.2 & 2.7 \\
		McDonald & 2459461.77778755 & 2.3 & 1.1 \\
		McDonald & 2459463.74475272 & -1.1 & 0.9 \\
		McDonald & 2459499.69725714 & 0.7 & 1.2 \\
		McDonald & 2459587.55677024 & -3.7 & 1.2 \\
		McDonald & 2459588.54027471 & -3.5 & 1.5 \\
		ATLAS & 2459599.16755742 & -4.3 & 1.7 \\
		TESS & 2459825.50977366 & -2.1 & 3.0 \\
		\hline
	\end{tabular}
\end{table}

\begin{figure*}
	\includegraphics[width=\textwidth]{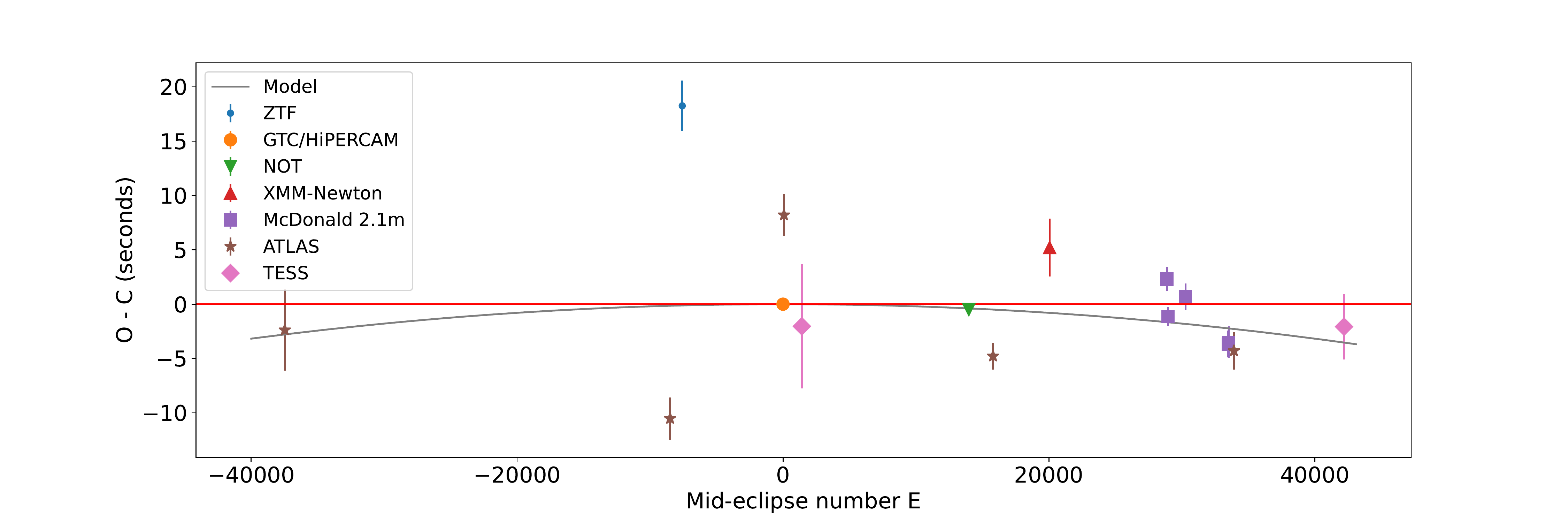}
    \caption{The theoretical O - C for ZTF J2130 obtained using $\dot P$ from \citet{kup20} (grey curve) shown along with the observed O - C values from our data.}
    \label{fig:o-c}
\end{figure*}


High cadence photometry data from HiPERCAM, NOT, McDonald and XMM-Newton were used as is, since all these were pointed observations. The light curves along with their LCURVE model fits and resulting residuals are shown in Fig. \ref{fig:hicam} - \ref{fig:xmm}. It is worth noting that the UV data from XMM-Newton were recorded as photon counts, and had relatively large residuals. Nevertheless, the observation was long enough for it to be still worthy of inclusion in timing analysis. 

For ZTF and TESS survey data, we selected only a subset of data with the purpose of optimising the number of data points to be high, but the total spread of data points to be short. This served the purpose of getting a good model fit while simultaneously making sure the period was practically constant over the duration of observations.

In case of the ATLAS survey data, the observations are too far spread out, and it is difficult to get even a few tens of data points without entering a timescale of months. In order to tackle this issue, we binned the points into groups of 200-300 to have enough data per group to fit a model. The challenge then was to have an initial value for a mid-eclipse time fitting. We compared the data with our calculated mid-eclipse times (C) and chose points corresponding to the closest observation times. As mentioned in the previous paragraph, it would be desirable to have data over a short span of time to ensure that the period is constant. For ATLAS however, we make an exception due to the large scatter over time.

We then use LCURVE with emcee \citep{for13} to implement an MCMC sampler that runs a number of parallel chains to converge at a solution. For large data sets (>1000 data points) we ran 128 parallel chains for 4000 generations, whereas for smaller data sets we ran 256 parallel chains for 4000 generations. The initial period was obtained using all available ZTF data (high cadence as well as scattered survey data). Recalling from earlier, the only free parameter while running LCURVE was $T_{0}$. We ensured that the solution converged clearly, with at least 2000 stable solutions at the end of every run. 

Following our first fits, we plotted a preliminary $O - C$ diagram. This initial plot showed a linear trend, likely corresponding to the linear term in equation~\ref{eq:o-c} ($\Delta P_{0}E$). This linear trend corresponds to a small offset of the true orbital period compared to our initial value from ZTF data. We fitted a line to this plot and used the slope to refine the period and remove the $\Delta P_{0}$ term. Subsequently, the light curve model fitting step was repeated with the refined period and new $T_{0}$ values were obtained. These values and their corresponding 1-$\sigma$ errors for all data sets are listed in Table~\ref{tab:o-c}. We also determined an updated ephemeris of -

\begin{equation}
    T_{0} \text{ (BJD)} = 2458672.68085911(8) + 0.0273195159(7) E
\end{equation}

The HiPERCAM light curve had the most precise $T_{0}$ fit, with a 1-$\sigma$ error of 0.1 seconds. Consequently, we used it as the reference mid-eclipse time $T_{0}$ and measured $O - C$ values for all other $T_{0}$ measurements accordingly. The final $O - C$ diagram was then plotted, as shown in Fig.\,\ref{fig:o-c}. The grey line shows $O - C$ values determined by substituting the refined period and $\dot P$ from \cite{kup20} in equation 1, with HiPERCAM mid-eclipse having $O - C = 0$. The scatter in $O - C$ values is relatively large and uneven with respect to the model. Although no clear orbital decay is apparent, we performed an RMS analysis to calculate an upper limit. We calculated the observed variability (RMS) of our $O - C$ values and demanded the variability due to orbital decay to be at least three times this value to assure a confident detection. This results in a corresponding $\dot P$ = $(-5.09) \times 10^{-12} s \text{ } s^{-1}$, placing an upper limit on the magnitude of orbital decay.

\section{Discussion and Conclusions}
\label{section:con}

We performed a high-resolution spectroscopic analysis of ZTF\,J2130 and ZTF\,J2055 in search for a direct detection of the accretion disk. The removal of the sdOB spectral lines and the phase-folding of the data led to a significant improvement in SNR but was still not sufficient to bring out any clear spectral features from the disk. We placed upper limits on the flux contribution by the disk to the total flux, obtaining a 3-$\sigma$ limit of 2.0\% for ZTF\,J2130 and 6.8\% for ZTF\,J2055. These limits are in agreement with predictions from theoretical models ($\sim$1\%) \citep{kup20}. It is important to note that the theoretical prediction of $\sim$1\% disk contribution is bolometric, and can be very different at different wavelengths. This prediction also depends on many assumptions made about the disk, and is therefore very uncertain.

Further improvement in our upper limits or perhaps a conclusive detection of signatures from the disk would require considerably more telescope time. Our observations of the two binaries were taken over two whole nights with the Keck telescope. Even doubling the SNR would demand a four-fold increase in telescope time, which is not practical. As a result, we have essentially exhausted the approach of optical spectral analysis with currently available state-of-the-art facilities.

The presence of an accretion disk could lead to a polarised signal with variations on the orbital period in the observed light. We do not detect any variations on our polarimetric observations with $4-\sigma$ limits of $\approx0.1\%$ in variability. The mean degree of polarisation is consistent with an origin in the interstellar medium. The non-detection of any polarimetric variability could be due to small contribution of the accretion disk to the total light of binary system. 

Neither the spectroscopic follow-up nor the polarimetric observations revealed any clear accretion disk signatures and as such the direct detection of an accretion disk in the ZTF\,J2130 and ZTF\,J2055 remains elusive. 

The short period of ZTF\,J2130 combined with the availability of data over more than six years motivated us to perform on $O - C$ analysis. The scatter of our $O - C$ values as well as errors for some of them were of the order or larger than the theoretically expected $O - C$ from the analysis in \citet{kup20}. Consequently, we have only placed an upper limit on the magnitude of the orbital decay parameter $\dot P$ based on our current set of observations. A major hurdle for this analysis was the use of data from several different telescopes. $O - C$ analysis is extremely sensitive to observation timestamps, and would require all telescope clocks to be perfectly synchronised with each other, and to be consistent with themselves over time. We believe that slight inconsistencies among telescope times could be one of the sources of errors. Additionally, small changes in the accretion rate or disk structure could lead to small changes in the light curve, potentially affecting timing measurements. The mass transfer could also be slowing down the orbital decay of this system. However, these effects would be difficult to quantify with currently available data. Nevertheless, we have reported an updated ephemeris using all our data. 

We conclude that it will take a few more years of observations to constrain a $\dot P$ observationally. We will continue to monitor ZTF\,J2130 with the McDonald 2.1 m telescope on regular intervals for this purpose.

\section*{Acknowledgements}

TK acknowledges support from the National Science Foundation through grant AST \#2107982, from NASA through grant 80NSSC22K0338 and from STScI through grant HST-GO-16659.002-A. SM acknowledges support from PRIN-MIUR 2017 UnIAM 2017LJ39LM.

This work has used observations obtained with the Samuel Oschin 48-inch Telescope at the Palomar Observatory as part of the Zwicky Transient Facility project. ZTF is supported by the National Science Foundation under Grant No. AST-1440341 and a collaboration including Caltech, IPAC, the Weizmann Institute for Science, the Oskar Klein Center at Stockholm University, the University of Maryland, the University of Washington, Deutsches Elektronen-Synchrotron and Humboldt University, Los Alamos National Laboratories, the TANGO Consortium of Taiwan, the University of Wisconsin at Milwaukee, and Lawrence Berkeley National Laboratories. Operations are conducted by COO, IPAC, and UW.

Some of the data presented herein were obtained at the W.M. Keck Observatory, which is operated as a scientific partnership among the California Institute of Technology, the University of California and the National Aeronautics and Space Administration. The Observatory was made possible by the generous financial support of the W.M. Keck Foundation. The authors wish to recognize and acknowledge the very significant cultural role and reverence that the summit of Mauna Kea has always had within the indigenous Hawaiian community. We are most fortunate to have the opportunity to conduct observations from this mountain. 

This research was supported in part by the National Science Foundation under Grant No. NSF PHY-1748958. This research also benefited from interactions that were supported by the Gordon and Betty Moore Foundation through grant GBMF5076.

Part of this work was based on observations made with the Nordic Optical Telescope, owned in collaboration by the University of Turku and Aarhus University, and operated jointly by Aarhus University, the University of Turku and the University of Oslo, representing Denmark, Finland and Norway, the University of Iceland and Stockholm University at the Observatorio del Roque de los Muchachos, La Palma, Spain, of the Instituto de Astrofisica de Canarias. DIPol-UF is a joint effort between University of Turku (Finland) and 
Leibniz Institute for Solar Physics (Germany).

This paper includes data collected by the TESS mission. Funding for the TESS mission is provided by the NASA's Science Mission Directorate.

We have also used data from the Asteroid Terrestrial-impact Last Alert System (ATLAS) project. ATLAS is primarily funded to search for near-Earth asteroids through NASA grants NN12AR55G, 80NSSC18K0284, and 80NSSC18K1575; byproducts of the NEA search include images and catalogues from the survey area. The ATLAS science products have been made possible through the contributions of the University of Hawaii Institute for Astronomy, the Queen’s University Belfast, the Space Telescope Science Institute, and the South African Astronomical Observatory.

This research made use of Astropy,\footnote{http://www.astropy.org} a community-developed core Python package for Astronomy \citep{astpy13, astpy18}.

\section*{Data Availability}

This work has made use of publicly available data from ZTF (\href{https://irsa.ipac.caltech.edu/Missions/ztf.html}{https://irsa.ipac.caltech.edu/Missions/ztf.html}), TESS (\href{https://archive.stsci.edu/missions-and-data/tess}{https://archive.stsci.edu/missions-and-data/tess}) and ATLAS (\href{https://fallingstar-data.com/forcedphot/}{https://fallingstar-data.com/forcedphot/}). Other data - photometric, polarimetric and spectroscopic - may be shared upon request to the authors.



\bibliographystyle{mnras}
\bibliography{refs} 








\bsp	
\label{lastpage}
\end{document}